\documentclass[12pt,twoside]{article}
\usepackage{a4}\usepackage{amsfonts}\usepackage{amsfonts}
\newcommand{\reals}{\mbox{${\rm I\!R }$}}
\newcommand{\be}{\begin{equation}}
\newcommand{\ee}{\end{equation}}
\newcommand{\bea}{\begin{eqnarray}}
\newcommand{\eea}{\end{eqnarray}}
\newcommand{\beq}{\begin{eqnarray}}
\newcommand{\eeq}{\end{eqnarray}}
\newcommand{\beao}{\begin{eqnarray*}}
\newcommand{\eeao}{\end{eqnarray*}}
\newcommand{\Ref}[1]{(\ref{#1})}
\newcommand{\res}{{\raisebox{-6pt}{$\begin{array}{c}{\rm \normalsize Res}\\ s=\frac32-n\end{array}$}}}
\newcommand{\resd}{{\raisebox{-6pt}{$\begin{array}{c}{\rm \normalsize Res}\\ s=\frac{D}2-n\end{array}$}}}

\newcommand{\bc}{boundary conditions }\newcommand{\bcp}{boundary conditions. }
\newcommand{\bck}{boundary conditions, }

\newcommand{\hkks}{heat kernel coefficients }
\newcommand{\hkksp}{heat kernel coefficients. }
\newcommand{\hkksk}{heat kernel coefficients, }
\newcommand{\hkk}{heat kernel coefficient }
\newcommand{\hkkp}{heat kernel coefficient. }
\newcommand{\mre}{multiple reflection expansion }
\newcommand{\mrep}{multiple reflection expansion. }
\newcommand{\mrek}{multiple reflection expansion, }
\newcommand{\nn}{\nonumber}
\newcommand{\pa}{\partial}
\newcommand{\al}{{\alpha}}
\newcommand{\la}{{\lambda}}
\renewcommand{\al}{{\alpha}}
\newcommand{\om}{{\omega}}
\newcommand{\Om}{{\Omega}}

\newcommand{\x}{{\vec{x}}}\newcommand{\y}{{\vec{y}}}

\newcommand{\all}{{\vec{a}}}
\newcommand{\pal}{\overleftarrow{\pa}}\newcommand{\pari}{\overrightarrow{\pa}}
\newcommand{\parl}{\stackrel{\leftrightarrow}{\pa}}

\begin{document}
\title{Multiple reflection  expansion and heat kernel coefficients}
\author{
  {\sc M. Bordag}\thanks{e-mail: Michael.Bordag@itp.uni-leipzig.de} \\
  {\sc D. Vassilevich}\thanks{e-mail: vassil@itp.uni-leipzig.de}\\
  \small  University of Leipzig, Institute for Theoretical Physics\\
  \small  Augustusplatz 10/11, 04109 Leipzig, Germany\\[6pt]
  {\sc H. Falomir}\thanks{e-mail:falomir@obelix.fisica.unlp.edu.ar}\\
  {\sc E.M. Santangelo}\thanks{e-mail:mariel@obelix.fisica.unlp.edu.ar}\\
  \small Departamento de F\'{\i}sica, Universidad Nacional de La Plata\\
  \small C.C.67, 1900 La Plata,Argentina
  }
\maketitle
\begin{abstract}
  We propose the multiple reflection expansion as a tool for the
  calculation of heat kernel coefficients. As an example, we give the
  coefficients for a sphere as a finite sum over reflections,
  obtaining as a byproduct a relation between the coefficients for
  Dirichlet and Neumann boundary conditions. Further, we calculate the
  heat kernel coefficients for the most general matching conditions on
  the surface of a sphere, including those cases corresponding to the
  presence of delta and delta prime background potentials. In the
  latter case, the multiple reflection expansion is shown to be
  non-convergent.
\end{abstract}
\thispagestyle{empty}
\section{Introduction}\label{Sec1}
Heat kernel coefficients play an important role in many areas of
theoretical physics. They govern the short-distance behavior of the
propagator and the small-time asymptotics of the Schr\"{o}dinger
equation. In quantum field theory, heat kernel coefficients define
the one-loop counter-terms and quantum anomalies, as well as the
large mass expansion of the effective action. It is clear,
therefore, that it is important to have an effective method of
calculation of these coefficients.

To the best of our knowledge,  heat kernel methods were first
applied to quantum physics by Fock in 1937 \cite{Fock}; then, they
were re-introduced by Schwinger in the 50's (see \cite{Schwinger}).
Due to De Witt \cite{DeWitt}, these methods became standard in
quantum field theory. The De Witt iteration procedure proved to
work  quite well on manifolds without boundaries and (after certain
improvements) allowed to calculate many terms in the asymptotic
expansion of the heat kernel
\cite{Amsterdamski:1989bt,Avramidi:1990ug,Avramidi:1991je,vandeVen:1997pf}.
On manifolds with boundaries, the methods based on functorial
properties of the heat kernel
\cite{Gilkey1975,Branson1990bt,Gilkey:1995} appeared to be more
appropriate. These methods allowed to calculate some higher terms of
the heat kernel expansion, e.g, for local boundary conditions
\cite{Branson:1999jz} and for a transmittal problem
\cite{Gilkey:2001mj}. The functorial methods being the most general
and the most powerful ones, they still have their limitations. They
work particularly well for most general operators in a certain
category. However, the number of independent invariants which can
enter a heat kernel coefficient grows very fast with the order of
the asymptotic expansion, so that combinatorics becomes unmanageable.
Alternatively, the general Seeley calculus, which is applicable for
general boundaries, may be used. But this method becomes unwieldy
beyond low orders.

The methods mentioned above are analytical, i.e., they produce local
formulae for the heat kernel coefficients in terms of the relevant
geometric invariants. An alternative to such methods are special case
calculations (see e.g.
\cite{Lyakhovsky:1991vv,Vassilevich:1991zi,Bordag:1996gm,Esposito1997,Kirsten:2000xc}
and references therein). In this case, the complexity of the
calculations is almost independent from the order of the asymptotic
expansion, but the method can only be applied to those problems
where a sufficiently high symmetry allows for the separation of
variables.

Another alternative is provided by iterative, resp., recursive
methods. Well known is the DeWitt iteration method. Less known are
re-formulations in terms of integral equations. For example, in
\cite{Bordag:1996fv}, the Lipmann-Schwinger equation for the
scattering problem was used to determine the asymptotic expansion of
the Jost function entering the regularized ground state energy.
While the known iterative methods work well for sufficiently smooth
background fields, an effective method working for singular
background fields or for boundary conditions is missing.

It is an aim of the present paper to suggest the \mre as such
method. In fact, it is based on an integral equation whose kernel is
located on the boundary. The iteration of this equation gives rise to
the \mrep The important point is that only a finite number of
reflections contribute to a given \hkkp

The use of the \mre in connection with vacuum energy is not new. In
\cite{bali70-60-401} it was employed to investigate the asymptotic
density of eigenvalues which, in the modern language, is equivalent
to the calculation of \hkksp In \cite{McAvity1991as} the possibility
of using the \mre was mentioned, but found to be too complicated for
a general boundary.   As far as we know, the method has never been
used as a tool for the calculation of the \hkksp In this paper, we
demonstrate its effectiveness with a recalculation of the \hkks on a
sphere. As a nice byproduct, we re-obtain a representation of the
coefficients as finite sums where the difference between Dirichlet
and Neumann \bc resides in certain signs only.

In general, the \mre can be viewed as some kind of perturbative
expansion. For instance, for imaginary frequencies it provides a well
convergent series for the propagator. It should be mentioned that,
in certain cases (for instance, with Dirichlet \bc), this is so
despite the absence of a small expansion parameter. So, the question
is whether this convergent behaviour is a general feature. The answer
is no, and we provide a counterexample by considering the most
general matching conditions on the surface of a sphere. They are
described by a four-parameter family and correspond, for instance, to
the presence of a delta function or its derivative on the surface. It
turns out that there is no expansion in powers of the parameter in
front of the derivative of the delta function but, instead, a nice
expansion in the inverse of this parameter, which cannot be at all
obtained by a \mrep

We would like to note that a Green function with a $\delta'$-function
potential has been considered before, for example, by path integral
methods in \cite{Grosche1995ah}. There, it was noticed that a
perturbative expansion similar to that for a $\delta$-function
potential yields some unsolvable relations, a fact that is not
surprising in view of the non-analyticity found by us.

The paper is organized as follows: In Sec. 2 we collect the
necessary formulas on spectral functions and their relations to the
\hkksp We write down the \mre in terms of integral equations for the
propagator as well as for the heat kernel. In Sec. 3 we use the \mre
in order to re-obtain the \hkk for the classical \bc on the sphere. In
Sec. 4 we consider the most general matching conditions on a sphere,
and calculate the corresponding \hkksp Sec. 5 contains a discussion
of our results and the conclusions. Some useful formulas are banned
into Appendix 1, while Appendix 2 contains the study of matching
conditions in higher dimensional spaces.

\section{Spectral functions}
In this section, we define the spectral functions to be
used in the rest of the present paper, and give a short introduction to perturbative
methods, supplemented with some examples of application.

Let us consider the Laplace operator on a domain
$\Om\in \mathbb{R}^{D}$. Let
$\Phi_{n}(\x)$ be its eigenfunctions, fulfilling Dirichlet or Neumann (or, more
generally, Robin) \bc on $S=\pa\Om$, $\la_{n}$ being the corresponding eigenvalues
:
\be\label{Lap}-\Delta \Phi_{n}(\x)=\la_{n}\Phi_{n}(\x).
\ee
(In Sec. \ref{singpotsec} we will consider the more complicated case
of matching conditions on a surface in $\mathbb{R}^{D}$). We
consider three local spectral functions. The first one is the
resolvent (propagator) $D_{\om}(\x,\y)$ (at imaginary frequency),
obeying the equation
\be\label{prop1}(\om^{2}-\Delta)D_{\om}(\x,\y)=\delta(\x-\y).
\ee
It can be represented as
\be\label{prop2}D_{\om}(\x,\y)=
\sum_{n}{\Phi_{n}(\x)\Phi_{n}(\y)\over\om^{2}+\la_{n}} .
\ee
The second spectral function is the zeta function, given by
\be\label{zeta1}\zeta(\x,\y;s)=\sum_{n}\la_{n}^{-s}{\Phi_{n}(\x)\Phi_{n}(\y)}
\ee
and, finally, the third is the heat kernel
\be\label{hk1}K(\x,\y|t)=
\sum_{n}{\Phi_{n}(\x)\Phi_{n}(\y)}e^{-t\la_{n}}.
\ee
These functions are related by means of
\bea\label{conn1}\Gamma(s)\zeta(\x,\y;s)&=&
\int_{0}^{\infty} dt \ t^{s-1}K(\x,\y|t)  \nn \\
&=&{2\over\Gamma(1-s)} \int_{0}^{\infty}d\om \ \om^{1-2s} D_{\om}(\x,\y).
\eea

In addition, we consider the corresponding global quantities, which appear as
integrals over $\Om$ of the local ones in the coincidence limit. Because of the distributional character of the \hkks
$a_{n}(\x,\x)$, it is
useful to introduce a test function $f(\x)$ into this
integration. So, let
\be\label{conn2}\zeta[f](s)=\int_{\Om}dx \ f(x)\zeta(\x,\x;s) ~~ \mbox{resp.}
~~ K[f](t)=\int_{\Om}dx \ f(x)K(\x,\x\mid t) \ee
be the global zeta function (resp. heat kernel). In many cases (as, e.g.,
for manifolds with boundaries and local boundary conditions)
the latter has an asymptotic expansion as $t\downarrow 0$
\be\label{exp2}K[f](t)\sim \ {1\over
  (4\pi t)^{D/2}}\sum_{n=0,\frac12,1,\dots}a_{n}[f] \  t^{n}\,.
\ee
We should warn the reader that the existence of the expansion
(\ref{exp2}) cannot be taken for granted.  For example, in the case
of some pseudo-differential operators or non-local boundary
conditions, $\ln t$ terms appear.

If the expansion (\ref{exp2}) exists, one can take $f=1$ to define the global
heat kernel coefficients:
\be\label{gcoeff}a_{n}=a_{n}[1]
\ee

If, apart from an appropriate behavior at large $t$, the heat kernel
has a power law asymptotics at small $t$, the zeta function
$\zeta[f](s)$ is a meromorphic function of $s$. From Eqs. \Ref{conn2}
and \Ref{exp2}, the coefficients $a_n$ can be represented by the
corresponding residua:
\be\label{res1}a_{n}[f]
=\resd(4\pi)^{D/2}\Gamma(s)\zeta[f](s)
~~~~(n=0,\frac12,1,\dots).
\ee
Furthermore, we remind the reader that, in general,
the coefficients consist of a bulk and a surface contribution
\be\label{coeff1}a_{n}[f]=\int_{\Om}dx f(x)b_{n}(\x)+\int_{\pa\Om}d\mu(\all)
f(\all)c_{n}(\all) ,
\ee
where we have used $\all\in\pa\Om$ as a notation for a point on the boundary,
as opposite to the genuine notation $\x\in\Om$ for a point in the bulk.

Now, we integrate Eq. \Ref{conn1} over the domain $\Om$ and insert the result
into Eq. \Ref{res1}. We thus arrive at
\be\label{anf}a_{n}[f]=\resd
 {2(4\pi)^{D/2}\over\Gamma(1-s)}\int_{0}^{\infty}d\om
\ \om^{1-2s}\int_{\Om}d\x \ f(\x) \ D_{\om}(\x,\x)
\ee
as the basic equation for calculating the coefficients out of the
propagator.

\bigskip

Having briefly reviewed some basic definitions and well-known facts, we
proceed now to a brief presentation of perturbative methods.

The perturbative expansion for the resolvent is constructed in the
following way: Let $D^0(\x ,\y )$ be a zeroth order resolvent.
Usually, $D^0$ is taken to be the free propagator in a flat space
without boundaries. Consider the Dyson equation
\begin{equation}
D(\x ,\y )=D^0(\x ,\y ) +\int_{\Sigma} d{\vec z} D^0(\x ,{\vec z})L
D({\vec z},\y ) \,, \label{Dyeq}
\end{equation}
where the integration goes over a sub-manifold $\Sigma \subset \Omega$, and
$L$ is some operator associated with the perturbation (see examples below).
Equation (\ref{Dyeq}) has the formal
solution:
\begin{equation}
D(\x ,\y )=D^0(\x ,\y )+\sum\limits_{n=1}^\infty
\int\limits_{\Sigma} d{\vec z_1}\dots
\int\limits_{\Sigma} d{\vec z_n} D^0 (\x ,{\vec z_1})LD^0({\vec z_1})\dots
LD^0({\vec z_n},\y )\,. \label{fsol}
\end{equation}
In \cite{Bordag:1999ed} it was shown that the heat kernel has a similar
representation,
\begin{eqnarray}
&&K(\x ,\y ;t)=K_0(\x ,\y ;t) + \sum_{n=1}^\infty (-1)^n
\int\limits_0^t ds_n \int\limits_0^{s_n}ds_{n-1} \dots \int\limits_0^{s_2}ds_1
\int\limits_{\partial M} d{\vec z_n}
\dots \int\limits_{\partial M} d{\vec z_1} \nonumber
\\
&& \qquad \times K_0(\x ,{\vec z_n};t-s_n)L\,
K_0({\vec z_n},{\vec z_{n-1}};s_n-s_{n-1})
\dots L\, K_0({\vec z_1},\y ;s_1), \label{solint}
\end{eqnarray}
where $K_0$ is a suitable chosen ``zeroth order'' heat kernel.

In order to clarify these definitions we consider, in what follows,
some examples.  Let $D^0=(-{\cal D})^{-1}$ be the propagator for a
second order differential operator ${\cal D}$. Let $\Sigma =\Omega$
and let $L$ be multiplication by a potential $V(\x )$.  This is the
standard situation of the DeWitt expansion with a smooth background
potential written in form of an iterated integral equation.  Then
the equations (\ref{Dyeq}) - (\ref{solint}) follow from the formal
expansion of $D=(-{\cal D}+V)^{-1}$ and $K=\exp (-t(-{\cal D}+V))$
respectively, so that $D(\x ,\y )$ and $K(\x ,\y |t )$ are the
propagator and the heat kernel of the operator $(-{\cal D}+V)$. If
the potential $V$ is smooth and falls off sufficiently fast at
infinity, all integrals in (\ref{solint}) exist. From dimensional
considerations, it is clear that the highest power of $V$ which may
contribute to the heat kernel coefficient $a_n$ is $V^n$. Therefore,
only the first $n$ terms of the expansion (\ref{solint}) must be
taken into account.

In our next example, let the operator $L$ be again the multiplication
by a potential $V$, but now, let $\Sigma$ be a subsurface of
co-dimension 1 in $\Omega$\footnote{This problem is a particular
case of a more general transmittal problem (see Sec.
\ref{singpotsec}).}. In \cite{Bordag:1999ed}, it was shown that all
terms in the expansion (\ref{solint}) exist and give a power-law
asymptotics of the heat kernel. Later, this expansion was used in
actual calculations of the heat kernel coefficients
\cite{Moss:2000gv}.

The Dyson equation is useful also for rather general perturbations
of boundary conditions as, e.g., for the case where more derivative
terms are added to the usual Neumann one (see \cite{Kummer:2000ae}).
In this case, however, dimensional arguments do not work and an
infinite number of terms contribute to any given heat kernel
coefficient $a_n$.

These examples demonstrate that the ``common sense'' arguments work
rather well. If there is a parameter $\epsilon$ in the theory such
that there is a smooth limit $\epsilon\to 0$ of the heat kernel
coefficients (such as $V\to 0$ above), then the formal expansions
(\ref{fsol}), (\ref{solint}) in that parameter usually give a good
approximation for the spectral functions. If such parameter is of
positive mass dimension, only a finite number of terms contribute to
each $a_n$.

This is, however, not the end of the story. In the next section we
will see that one can construct a perturbative expansion, the so
called multiple reflection expansion, even when no parameter or
limiting procedure exist. Moreover, also in this case, only a finite
number of terms contribute to each heat kernel coefficient - a
result which is hard to predict on the basis of ``common sense''
arguments.

\section{Multiple reflection expansion applied to the \hkks for the sphere}
This section contains a short overview of one particular perturbative method,
which is particularly well-suited for the treatment of boundary problems,
i.e., the multiple reflection expansion for Dirichlet and Robin boundary
conditions. Balian and Bloch \cite{bali70-60-401} applied this expansion, in
the boundary value problem context, to calculate the density of eigenvalues,
which is related to the heat kernel by a simple integral transformation. In
their work \cite{bali78-112-165}, they pointed out that the divergent part of
the Casimir energy is given by a few first reflection contributions.  This
fact, however, has not been fully appreciated. Therefore, we find it useful to
repeat some basic facts, translating them to a more modern language, and to
supply the reader with a simple example. In doing so, we omit many details
which can be found in the original literature
\cite{bali70-60-401,Hansson:1983xt}.

The \mre is based on simple formulas known from electrostatics: Having in mind the application to Dirichlet
boundary conditions, let $\mu(\all)$
be the density of a double layer (dipole layer) on a surface $S$. The
corresponding potential is
\be\label{dl}\Phi(\x)=
\int_{S}d^2a_1 \ \Delta_{\om}(\x-\all_{1}) \ \pal_{\all_{1}} \
\mu(\all_{1}),
\ee
where $\pal_{\all_{1}}$ is the normal derivative, restricted to the surface
$S$ and acting to the left. Explicitly written, it reads
$\Delta_{\om}(\x-\all) \
\pal_{\all}=\vec{n}(y)\vec{\nabla}_{y}\Delta_{\om}(\x-\y)_{|_{\y=\all}}$,
where $\vec{n}$ is the normal vector. The measure on $S$ is $d^2a_1=d
u_{1}d u_{2}\sqrt{g}$ where $(u_{1},u_{2})$ are the coordinates of a point
$\all(u_{1},u_{2})$ on $S$ and $g_{ij}={\pa\all\over\pa u_{i}}{\pa\all\over\pa
  u_{j}}$ is the metric. In Eq. \Ref{dl}, the propagator $\Delta_{\om}(\x-\y)$
is the free one, i.e., without \bcp In three dimensions it is simply the
Yukawa potential
\be\label{free}\Delta_{\om}(\x-\y)
={e^{-\om r}\over 4\pi \ r} ~~~~~~~~~ (r=|\x-\y|).
\ee
The potential $\Phi(\x)$ is discontinuous for $\x$ approaching the surface
$S$ ($\x\to\all$) and the equation
\be\label{dlrel}lim_{\x\to\all}\Phi(\x)=
\int_{S}d^2a_1 \ \Delta_{\om}(\all-\all_{1}) \ \pal_{\all_{1}} \
\mu(\all_{1}) +\frac12 \mu(\al)
\ee
holds. The additional
contribution (last term) appears due to the fact that limit and integration do not commute.

In a similar fashion, having in mind the application to
Neumann \bck the potential $\chi(\x)$ of a charged surface with charge density
$\rho(\all)$
\be\label{sl}\chi(\x) = \int_{S}d^2a_1
\ \Delta_{\om}(\x-\all_{1}) \ \rho(\all_{1})
\ee
has a discontinuous derivative,
\be\label{slrel}\lim_{\x\to\all} \ \vec{n}\vec{\nabla}_{x} \ \chi(\x)=
\int_{S}d^2a_1 \ \pari_{\all} \ \Delta_{\om}(\all-\all_{1}) \
\rho(\all_{1}) -\frac12 \rho(\al).
\ee

In general, the \mre for the resolvent reads
\bea\label{mre1}D_{\om}(\x,\y)&=&\Delta_{\om}(\x-\y)+\kappa
\int_{S}d^2a_1\Delta_{\om}(\x-\all_{1})\parl_{\all_{1}}\Delta_{\om}(\all_{1}-\y)   \\
&&+\kappa^{2}\int_{S}d^2a_1\int_{S}d^2a_2
\Delta_{\om}(\x-\all_{1})\parl_{\all_{1}}\Delta_{\om}(\all_{1}-\all_{2})
                         \parl_{\all_{2}}\Delta_{\om}(\all_{2}-\y)  \nn \\ &&
+\dots      \nn
\eea
with the notation $\parl=\pal+\pari$. For $\kappa=1$ this propagator obeys
Dirichlet and, for $\kappa=-1$, Neumann \bcp The validity of this expression can be verified by noting that
it fulfills the differential equation for $\x\not\in S$. Moreover, \bc can be
checked using \Ref{dlrel} and \Ref{slrel}, whereby the additional contributions
give rise to cancellations between successive orders of
reflections. The expansion \Ref{mre1} is called \mre because it can be
interpreted as a motion described by the free propagator from $\x$ to
$\all_{1}$, being reflected (however under any angle due to the integration
over $\all_{1}$), further moving to $\all_{2}$ and so forth.   More details
can be found in \cite{Hansson:1983xt} and related papers\footnote{It must be
stressed that despite its simple form, the derivation of equation
(\ref{mre1}) contains several subtle points, which are explained in the
Appendix of \cite{Hansson:1983xt}.}.

A simple example for the \mre appears if the surface $S$ is a sphere. In this
case, the expansion becomes an algebraic one. It can be obtained from
(\ref{mre1}) by turning to spherical coordinates. It is,
however, easier to use the known expression for the exact propagator with given boundary conditions,
\be\label{kprop1}D_{\om}(\x,\x')=\sum_{l,m}Y_{l,m}(\theta,\varphi)Y_{l,m}^{*}(\theta',\varphi')
D_{l}(r,r') ,
\ee
with
\be\label{kprop2}D_{l}(r,r')={1\over\sqrt{rr'}}\left(I_{\nu}(\om
  r_{<})K_{\nu}(\om r_{>})-I_{\nu}(\om r)I_{\nu}(\om r')K_{D,R}\right)
\ee
and $\nu\equiv l+\frac12$. Here, $I_{\nu}(x)$ and $K_{\nu}(x)$ are the
modified Bessel functions, and we have introduced the notations
\be\label{KD}K_{D}={K_{\nu}(\om R)\over I_{\nu}(\om R)} ,
\ee
for Dirichlet \bc, and
\be\label{KR}K_{R}={{\pa\over\pa R}\left( R^{u}K_{\nu}(\om R)\right)
\over {\pa\over\pa R}\left( R^{u}I_{\nu}(\om R)\right)}
\ee
for Robin boundary conditions, where the solutions of Eq. \Ref{Lap} have to
fulfill $ {\pa\over\pa r}\left(r^{u+\frac12}\phi_{n}(r)\right)_{|_{r=R}}=0$.
For $u=-\frac12$, these reduce to Neumann \bc on the 2-dimensional sphere.

The \mre appears in the following way \cite{Hansson:1983xt}. Represent
\be\label{1655}K_{D}={K_{\nu}(\om R)K'_{\nu}(\om R)\over
I_{\nu}(\om R)K'_{\nu}(\om R)},
\ee
and use the Wronskian
$I'_{\nu}(x)K_{\nu}(x)-I_{\nu}(x)K'_{\nu}(x)=1/x$
to rewrite the denominator in (\ref{1655}) as
\be\label{}I_{\nu}(\om R)K'_{\nu}(\om R)={-1\over 2\om R}\left(1-\om
  R{\pa\over\pa\om R}\left(I_{\nu}(\om R)K_{\nu}(\om R)\right)\right) .
\ee

Next, expand this denominator so that one obtains for $K_D$ the representation
\be\label{KDexp}K_{D}=-2\om R K_{\nu}(\om R)K'_{\nu}(\om R)
\sum_{k=0}^{\infty}\left(\om
  R{\pa\over\pa\om R}\left(I_{\nu}(\om R)K_{\nu}(\om R)\right)\right)^{k}.
\ee
In a similar way, one obtains
\bea\label{KRexp}K_{R}&=&-{2\om R } K_{\nu}(\om R)\left({ K'_{\nu}(\om
  R)}+\frac{u}{\om R}K(\om R)\right)  \\
&&  \times  ~ \sum_{k=0}^{\infty}(-1)^{k}\left(
{\om R}
\left({\pa \left(I_{\nu}(\om R)K_{\nu}(\om R)\right)\over \pa \om R}
+\frac{2u}{\om R} I_{\nu}(\om R)K_{\nu}(\om R)\right)
\right)^{k}. \nn
\eea
This formal expansion has been shown \cite{Hansson:1983xt}
to be equivalent to the \mre
\Ref{mre1}, where the number of reflections is $k+1$.

In view of Eq. \Ref{anf}, we perform the integration over the surface of the
sphere and define
\be\label{kprop3}D_{\om}(r)=\int_{\pa\Om}d\mu(\all)D_{\om}(\x,\x)=
\sum_{l=0}^{\infty}(2l+1)D_{l}(r,r)
\ee
so that the coefficients $a_{n}$, Eq. \Ref{gcoeff}, turn out to be given by
\be\label{and}a_{n}=\res {16\pi^{3/2}\over\Gamma(1-s)} \int_{0}^{R}dr \
r^{2}\sum_{l=0}^{\infty}(2l+1) \int_{0}^{\infty}d\om \ \om^{1-2s}D_{l}(r,r).
\ee

The procedure to calculate the coefficients from this representation is as
follows: First, we remark that the poles in $s$ result from large, both $\om$
and $l$, in the Bessel functions. The poles corresponding to boundary
contributions (the $c_{n}$ in Eq. \Ref{coeff1}) appear, in addition, from the
upper limit of the integration over $r$. So, we use the uniform asymptotic
expansion of the Bessel functions (it is given in the Appendix), together with
the \mre \Ref{KDexp} or \Ref{KRexp}, and insert them into Eq. \Ref{and}.

Let us start with the first contribution in the rhs. of Eq.
\Ref{kprop2}. It corresponds to the free space propagator and, thus,
it doesn't know about the boundary. Consequently, it gives the volume
contribution, which is $a_{0}=\frac{4\pi}{3}R^{3}$.

In order to calculate the higher coefficients, we consider the second term in
the rhs. of Eq. \Ref{kprop2}. According to the sum in the rhs. of Eq.
\Ref{KDexp} resp. Eq. \Ref{KRexp} we represent the coefficients as a sum over
reflections,
\be\label{splitan}a_{n}=\sum_{k=0}^{2n}a_{n}^{(k)}. \ee
Using the uniform asymptotic expansion of the Bessel functions these
coefficients can be calculated (for details, see Appendix 1).  As a
result, for Dirichlet boundary conditions, the first $a_{n}^{(k)}$'s
are:

\bigskip

\begin{tabular}{c|cccccc|ccccccccccccccccccc}
    k=     &0&  1   &  2    &3&4&5           &  $a_{n}$             \\[3pt]
    \hline
$n=\frac12$  &$ {-2\pi^{3/2}}$   &&&&&&$ {-2\pi^{3/2}}$               \\[4pt]
$n=1$       &  $2\pi $  &  $\frac23 \pi $  &&&&& $\frac83 \pi $ \\[4pt]
$n=\frac32$  &  0  & 0&  $-\frac16 \pi^{3/2} $ &&&&$-\frac16 \pi^{3/2} $                 \\[4pt]
$n=2$       &0&   $-\frac{4}{35}\pi $  &$-\frac{1}{21}\pi $
&$-\frac{1}{9}\pi $  &  && $-\frac{16}{315}\pi $ \\[4pt]
$n=\frac52  $&  0&0&0&$ \frac{1}{80}\pi^{3/2}$&$
-\frac{1}{48}\pi^{3/2}$&$ \frac{1}{12}\pi^{3/2}$&$ -\frac{1}{120}\pi^{3/2}$\\[4pt]
$n=3$& 0& $-\frac{40}{3003}\pi $& $-\frac{2}{143}\pi $& $\frac{12}{715}\pi $& $-\frac{1}{130}\pi $& $\frac{1}{90}\pi $& $-\frac{64}{9009}\pi $
\end{tabular}.\\[12pt]

The coefficients $a_{n}$ are proportional to $R^{3-2n}$. In this
table we have taken $R=1$.  When replaced in Eq.  \Ref{splitan},
they sum up to the known values (shown in the last column) as can be
checked, for example, by comparing them with the results in appendix
B of \cite{Bordag:1996gm}). It is interesting to note that the
corresponding \hkks for Robin \bc with $u=0$ can be obtained through
the replacement $a_{n}^{(k)}\to(-1)^{k}a_{n}^{(k)}$.

As a last example in this section, we give some reflection contributions to
the \hkk $a_{2}$ for Robin \bcp They read ($R=1$)
\beao\label{a2R}a_{2}^{(0)}&=&0,  \\
a_{2}^{(1)}&=& \frac{4 \pi}{105}(-3 + 12u - 28u^2)  ,   \\
a_{2}^{(2)}&=&  \frac{ \pi}{105}  (5 - 42u+ 140u^2 - 280u^3), \\
a_{2}^{(3)}&=&  \frac{ \pi}{315} (35 - 270u + 756u^2 - 840u^3). \eeao
The sum \Ref{splitan} gives the known result
$a_{2}=\frac{2\pi}{45R}(1 - 18 u + 60 u^{2}  - 120 u^{3}) $. In
particular, Neumann \bc on a sphere  follow by choosing $u=-\frac12$.


\section{Singular potentials on a spherical shell}\label{singpotsec}

In this section we will study the heat kernel expansion for the free
Laplacian in ${\reals}^D$, acting on the space of functions obeying
on a $D-1$-dimensional sphere, $S^{D-1}$, certain matching conditions
relating the values of the functions and their first derivatives on
different sides of the sphere. If one assumes that the matching
conditions are ultra-local in angular coordinates (they do not
contain tangential derivatives), the most general choice is the
following four-parameter family \cite{Albeverio88}:
\begin{eqnarray}
&&\phi_+=\omega a\phi_-+\omega b\phi_-' \,,\nonumber \\
&&\phi_+'=\omega c \phi_-+\omega d\phi_-' , \label{gmc}
\end{eqnarray}
where
\begin{equation}
\phi_\pm =\lim_{r\to R\pm 0} \phi (r)\,,\quad
\phi_\pm' =\lim_{r\to R\pm 0} \partial_r \phi (r)\,.
\label{defpm}
\end{equation}
Here, $\omega$ is a complex phase factor, which we include for
completeness only. We consider real fields and put $\omega=1$. The
other parameters obey the restriction $ad-bc=1$.

There are two important special cases of the conditions (\ref{gmc}). Take
\begin{equation}
a=d= 1,\quad b=0. \label{dirlim}
\end{equation}
This requires the functions to be continuous across the surface and
their derivatives to have a jump. This is equivalent to having a
delta function potential $V(x)=c\delta(r-R)$ on $S^{D-1}$, which can
be viewed as a singular background potential concentrated on the
surface. The formal limit $c\to\infty$ turns this matching condition
into Dirichlet boundary conditions ($\phi_\pm=0$).

The other special case is
\begin{equation}
 a=d=1, \quad c=0,\label{roblim}
\end{equation}
requiring the derivatives to be continuous, and the functions
themselves to have a jump. This is usually attributed to the
presence of a background potential in the form of the derivative of
the delta function. The formal limit $b\to\infty$ turns this
condition into Robin boundary conditions ($\frac{a}{b}
\phi_\pm+\phi_\pm'=0$).

In general, the parameters $a,b,c,d,\omega$ may depend on the angular
coordinates on $S^{D-1}$.  In this paper we restrict ourselves to
the case where there is not such dependence. Then, variables can be
easily separated by making the ansatz
\begin{equation}
\phi_{(n)}(x)= r^{\frac {2-D}2} \phi_{n,l}(r)Y_{(l)}(\Omega )\,,
\label{separ}
\end{equation}
where $Y_{(l)}(\Omega )$ are spherical harmonics depending on the
angular coordinates $\Omega$. Once such ansatz is adopted, the radial
functions $\phi_{n,l}$ must satisfy the equation
\begin{equation}
\left[ \frac{d^2}{dr^2} +\frac 1r \frac d{dr}
-\frac{\nu^2}{r^2} +\lambda^2_{n,l} \right]
\phi_{n,l}=0 \label{radeq}
\end{equation}
with $\nu =l+\frac{D-2}2$, and the matching conditions (\ref{gmc}) at $r=R$
with shifted values of the constants:
\begin{equation}
a\to\bar a=a+\frac{2-D}{2R} b \,,\qquad
c\to\bar c=c+\frac{2-D}{2R} d \,.\label{barac}
\end{equation}

The degeneracy of each eigenvalue $\lambda^2_{n,l}$ is
\begin{equation}
d_l(D)=\frac{(2l+D-2)(l+D-3)!}{l!(D-2)!} \,.\label{degen}
\end{equation}

In what follows, we will determine the corresponding zeta function
and, from it, the corresponding heat kernel coefficients. Because we
have a continuous spectrum, we must separate the translational
invariant part (it does not depend on the background). We use the
procedure described in Ref. \cite{Bordag:1996fv} using the setup of a
scattering off the background potential. We have to define the
so-called regular solutions $\phi_{p,l}$ which have the same
behavior at $r\to 0$ as the free solution
\begin{equation}
\phi_{p,l}(r)\sim J_\nu (pr) .\label{regsol}
\end{equation}
The behavior of this regular solution for $r\to\infty$ defines the Jost
function $f_l(p)$:
\begin{equation}
\phi_{p,l}(r)= f_l(p)H^{(2)}(pr)
+f^*_l(p)H^{(1)}(pr) \ .\label{defJost}
\end{equation}

In the present case, the eigenfunctions of the Laplace operator can be found
exactly and they give, for the problem at hand,  the Jost function
\begin{eqnarray}\label{freal}
f(p)&=& \frac{\pi pR}{4i} \Big[
a J_\nu (pR)H^{(1)}_\nu (pR)' +
b p J_\nu (pR)' H^{(1)}_\nu (pR)'
\\
& & ~~~~~~~  -\frac{\bar c}{p} J_\nu (pR)H^{(1)}_\nu (pR) -
          d J_\nu  (pR)' H^{(1)}_\nu (pR) \Big] \,.       \nn
\end{eqnarray}

Now, in order to use a formula like Eq. \Ref{zeta1} we need to have discrete
eigenvalues. So we suppose for a moment that our system is placed inside a
sphere of larger radius $R^{*}$. Imposing Dirichlet boundary conditions at
$r=R^{*}$, we obtain the following equation for the eigenvalues
$p=\lambda_{n,l}$:
\begin{equation}
f_l(p)H^{(2)}(pR^{*} )+f^*_l(p)H^{(1)}(pR^{*} )=0 \,. \label{zeroatrho}
\end{equation}
Then, the $\zeta$-function can be represented as a contour integral:
\begin{equation}
\zeta (s)=\sum\limits_{l=0}^\infty d_l(D) \int\limits_\gamma
\frac {dp}{2\pi i} (p^2+m^2)^{-s} \frac {\partial}{\partial p} \ln
\left[ f_l(p)H^{(2)}(pR^{*} )+f^*_l(p)H^{(1)}(pR^{*} )\right]
 \,.\label{int+sum}
\end{equation}
The contour $\gamma$ is chosen counterclockwise, enclosing all
solutions of Eq. \Ref{zeroatrho} on the positive real semi-axis and
the positive imaginary semi-axis. For convenience, we have
introduced an auxiliary mass, which we will later put to zero. There
is a cut in the complex plane, which goes from $im$ to $i\infty$.
Since the number of negative modes of the Laplacian is finite, we
can always choose $m$ to be sufficiently large so that all poles of
the integrand (\ref{int+sum}) are below $im$.  Next, we may deform
the integration contour as described in
\cite{Bordag:1996gm,Bordag:1996fv,Kirsten:2000xc} to go along the
two sides of the cut. We perform the limit $R^{*}\to\infty$, and drop
some contributions which are exponentially small in this limit and a
term which does not depend on the matching conditions (i.e. the
``empty space'' contribution).  The procedure sketched above is a
quite general one, and not specific of this example, since it uses
only some general properties of the scattering problem, as
Hermiticity and ellipticity of the Laplacian.

Next, we take the limit
$m\to 0$, which is smooth at least for the heat kernel
asymptotics, and obtain:
\begin{equation}
\zeta (s)=\frac {\sin (\pi s)}{\pi}
\sum\limits_{l=0}^\infty d_l(D)\int\limits_0^\infty dk \
k^{-2s} \partial_k (\ln f_l(ik)) \,.\label{imk}
\end{equation}
In the Jost function, we can drop any constant factor since it does not contribute to (\ref{imk}) and redefine
\begin{equation}
f_l(ik)=1+\gamma k(IK)' +\beta k^2 I'K' +\alpha IK \,\label{Jostabc}
\end{equation}
with new parameters $\gamma=(d-\bar{a})R/(d+\bar{a})$,
$\beta=-2bR/(d+\bar{a})$ and $\alpha=2\bar{c}R/(d+\bar{a})$ as well as the
short-hand notations $I=I_{\nu}(kR)$, $K=K_\nu (kR)$.

In order to get the poles of the zeta function \Ref{imk}, thus
determining the \hkks by means of Eq. \Ref{res1}, we insert into
this Jost function the uniform asymptotic expansion, Eq. \Ref{uae},
of the modified Bessel functions and obtain
\be  \label{aa}\ln f_{l}(ik)= \ln
\left( 1
-\beta  \frac{\nu}{2t} i^{d}_{\nu}(t)k^{d}_{\nu}(t)
+\alpha \frac{t}{2\nu} i_{\nu}(t)    k_{\nu}(t)
+\frac{\gamma}{2}     (i^{d}_{\nu}(t)k_{\nu}(t)-i_{\nu}(t)k^{d}_{\nu}(t))
               \right).
\ee Now, because all functions,
$i_{\nu}(t),i^{d}_{\nu}(t),k_{\nu}(t)k^{d}_{\nu}(t)=1+O\left(\frac1\nu
\right) $ are of order one for $\nu\to\infty$, the leading
contribution in the argument of the logarithm is the one proportional
to $\beta$. As this term grows with $\nu$, two cases must be treated
separately, i.e., $\beta=0$ and $\beta\ne0$.

For $\beta=0$ we obtain, by means of Eqs. \Ref{uae1}, an expansion
similar to Eq. \Ref{Ykpi}, where the $Y_{kpi}$ are polynomials in the
coefficients $\alpha$ and $\gamma$. The remaining calculations run
in the same manner as in the preceding section and we obtain, in
$D=3$ dimensions,
\bea \label{bn3}
a_{1}&=&-4\,\left( \alpha - \gamma \right) \,\pi  \\
a_{3/2}&=&{\left( \alpha - \gamma \right) }^2\,{\pi }^{\frac{3}{2}} \nn \\
a_{2}&=&- \frac{2}{15}\,\left( 5\,{\alpha}^3 - 120\,\gamma - 5\,{\alpha}^2\,
     \gamma + 3\,\alpha\,{\gamma}^2 - 3\,{\gamma}^3 \right) \,\pi \nn \\
a_{5/2}&=& \frac{1}{8}\left( {\alpha}^4 - 20\,\alpha\,\gamma - 2\,{\alpha}^3\,
      \gamma +  36\,{\gamma}^2 + 2\,{\alpha}^2\,{\gamma}^2 -
    2\,\alpha\,{\gamma}^3 +    {\gamma}^4 \right) \,{\pi }^{\frac{3}{2}}.\nn
\eea
The corresponding results for higher dimensions are given in
Appendix 2.

Next, we turn to the case $\beta\ne 0$, which corresponds to the
presence of a $\delta^{'}$ potential. Here, we rewrite the logarithm
of the Jost function \Ref{Jostabc} in the form \bea\label{bet}
  \ln f_{l}(ik)&=&\ln \beta +\ln\frac{\nu}{2t}\\
  && +\ln
  \left(1+\left(i^{d}_{\nu}k^{d}_{\nu}-1\right)-\frac{2t}{\beta\nu}-\frac{\alpha}{\beta}\frac{t^{2}}{\nu^{2}}
    i_{\nu}k_{\nu}-\frac{\gamma}{\beta}\frac{t}{\nu}\left(i^{d}_{\nu}k_{\nu}-i_{\nu}k^{d}_{\nu}\right)
  \right) \,.\nn \eea The first term in the rhs, $\ln\beta$, drops out due to the
  derivative in \Ref{imk}. The contributions surviving in the limit
  $\beta\to\infty$ in \Ref{bet} are just the same as those one obtains for Neumann
  \bcp Inserting now the asymptotic expressions \Ref{uae1} and proceeding as
  above, one arrives at the following coefficients
in $D=3$ dimensions,
\begin{eqnarray}
a_{1}  &=&16\frac{1}{\beta} \ {\pi }\nn\\
a_{3/2}&=& \frac{1}{3}\left(1+24 \frac{\alpha}{\beta}+16\frac{1}{\beta^{2}}
     \right) \,{\pi }^{\frac{3}{2}} \nn\\
a_{2}&=&\frac{8}{15}\left( \frac{3}{\beta}
    +60\,\frac{\alpha}{\beta^{2}}
        + 20\frac{\gamma}{\beta^{2}}+ 80\frac{1}{\beta^{3}} \right) \ \pi\nn\\
a_{5/2}&=&\frac{1}{30}
\left( 2
      +  40\,\frac{\alpha}{\beta}
       + 15\,\frac{\gamma}{\beta}
      + \frac{80}{\beta^{2}}
       +  120\,\frac{\alpha^2}{\beta^{2}}
       + 120\,\frac{\alpha\gamma}{\beta^{2}}
         + 40\frac{\gamma^2}{\beta^{2}}
    \right.  \nn\\   &&\qquad\qquad\qquad      \left.
      +  960\,\frac{\alpha}{\beta^{3}}
       + 480\frac{\gamma}{\beta^{3}}
      +  \frac{960}{\beta^{4}}
  \right) \pi^{\frac 32 }\label{D3}  .
\end{eqnarray}
Again, the corresponding results for higher dimension are given in
Appendix 2. As already pointed out in the Introduction, the
coefficients present, in this case, a dependence on inverse powers
of $\beta$.

\section{Conclusions}
In the foregoing sections we used the \mre as a method for the
calculation of \hkkp As a simple example, we considered \bc on a
sphere and obtained the \hkks as a finite sum over reflections, Eq.
\Ref{splitan}. An interesting point is a connection between
Dirichlet and Neumann \bc following from this representation: The
contributions from the reflections are the same in both cases except
for the sign for odd number of reflections.  This can be clearly
seen already from Eq. \Ref{mre1} and, hence, holds in general. It
occurs that this seemingly simple observation had not been spelled
out before.

The \mrek as well as the equivalent integral equations, Eq.
\Ref{fsol} for the propagator and Eq. \Ref{solint} for the heat
kernel, provide a perturbative expansion. For the propagator, this
expansion is convergent for imaginary frequency (as used in this
paper), as was already observed in \cite{bali70-60-401}. For real
frequencies it may diverge. The same holds, presumably, for the heat
kernel: The corresponding perturbative expansion can be expected to
converge. It is
interesting to note that the convergence of these expansions does
not follow from a small expansion parameter. For instance, with
Dirichlet \bc there is no such parameter whereas for matching
conditions corresponding to a delta function potential on the
surface there is one, cf. \cite{Bordag:1999ed} and Sec. 4.  The
corresponding quantities in the expansion may be numbers which turn
out to be sufficiently small.  In the example with Dirichlet \bck in
Eq. \Ref{KDexp}
\be\label{}\om
  R{\pa\over\pa\om R}\left(I_{\nu}(\om R)K_{\nu}(\om R)\right) <1
\ee
holds ensuring the convergence of the geometric series there.

In general, the convergence issue is not trivial. As an example, we
considered in Sec. 4 the most general background potential
concentrated on a spherical surface. It is given by the matching
conditions in Eq. \Ref{gmc}, which include a delta function potential
and its derivative as special cases (Eqs. \Ref{dirlim} and
\Ref{roblim}). Using the techniques introduced in
\cite{Bordag:1996gm}, we calculated for the first time the
corresponding \hkksp The lesson with respect to the \mre is that,
for $\beta\ne0$ (in Eq. \Ref{Jostabc} (or, equivalently, for $b\ne0$
in (\Ref{gmc})), i.e., in the presence of the derivative of the delta
function, the coefficients are not analytic in $\beta$.  In fact,
they are polynomials in $\frac1\beta$. Hence, the \mre cannot
converge for $\beta\ne0$.

To summarize, we have stressed the convenience of using the \mre for
the calculation of \hkksk while showing, at the same time, some
limitations of the method. In general, this method provides, after
the general Seeley's calculus, the only systematic way to calculate
\hkks for manifolds with boundary, and we expect that it will be
useful in future applications.
\section*{Acknowledgment}
We thank K. Kirsten for helpful discussions. The work was supported
by Antorchas-DAAD (grant A-13740/1-85) and by the DFG (grant Bo
1112/11-1).
\section*{Appendix 1}
In order to calculate the coefficients $a_{n}^{(k)}$ for $n\ge\frac12$ in Eq.
\Ref{splitan} it is useful to carry out the integration over $r$ in
(\ref{and}) using the known formula $\int dx \ x \ I_{\nu}(x)^{2}= $ \break
$\frac{x^{2}}{2}\left(I_{\nu}(x)^{2}\left(1+\frac{\nu^{2}}{x^{2}}\right)-I'_{\nu}(x)^{2}\right)$.
Inserting the second term in the rhs. of Eq. \Ref{kprop2} results in the
representation
\be\label{ank}a_{n}=\resd
{16\pi^{3/2}R^{2s}\over\Gamma(1-s)}
\sum_{l=0}^{\infty}\nu^{3-2s}
\int_{0}^{\infty}dz \ z^{1-2s}
\left(I^{2}\left(1+\frac{1}{z^{2}}\right)-{I'}^{2}\right)
K_{D,R} \,,
\ee
where  $I\equiv I_{\nu}(\nu z)$, $K\equiv K_{\nu}(\nu z)$
and we introduced a new variable $z=\frac{\om R}{\nu}$.

Next, we substitute the uniform asymptotic
expansions of the modified Bessel functions for
$\nu\to\infty$, $z$ fixed:
\be\label{uae}
\begin{tabular}{rclrcl}$
I_{\nu}(\nu z)$&=&$
{1\over\sqrt{2\pi\nu}}{e^{\nu\eta}\over(1+z^{2})^{1/4}} \ i_{\nu}(t),$
&
$ K_{\nu}(\nu z)$&=&
$\sqrt{{\pi\over 2\nu}}{e^{-\nu\eta}\over(1+z^{2})^{1/4}} \ k_{\nu}(t),$    \\
$I'_{\nu}(\nu z)$&=&
${e^{\nu\eta}\over\sqrt{2\pi\nu}}{(1+z^{2})^{1/4}\over z} i^{d}_{\nu}(t),  $  &$
K'_{\nu}(\nu z)$&=&
$-\sqrt{{\pi\over 2\nu}}{(1+z^{2})^{1/4}\over z}e^{-\nu\eta} \ k^{d}_{\nu}(t) ,   $
\end{tabular}
\ee
with
\bea \label{uae1}
    i_{\nu}(t)= \sum\limits_{r\ge 0}        {u_{r}(t)\over\nu^{r}}, \ &&
k^{d}_{\nu}(t)= \sum\limits_{r\ge 0}{(-1)^{r}u_{r}(t)\over\nu^{r}}, \ \\\nn
    i_{\nu}(t)= \sum\limits_{r\ge 0}        {v_{r}(t)\over\nu^{r}}, \ &&
k^{d}_{\nu}(t)= \sum\limits_{r\ge 0}{(-1)^{r}v_{r}(t)\over\nu^{r}} .
\eea
Here, the notation $t=1/\sqrt{1+z^{2}}$ is used. The Debye polynomials
$u_{r}(t)$ and $v_{r}(t)$ can be found in \cite{abra70b}, they contain powers
of $t$ from $r$ to $3r$. We don´t need the function $\eta$, since it cancels
out in our case.

We can thus write
\be\label{Ykpi}\left(I^{2}(1+\frac{1}{z^{2}})-I'^{2}\right)K_{D,R} =
{\mu^{k}2^{-k}\over 2\nu z^{2}} \sum_{k=0}^{2n}
\sum_{p=k+1}^{2n+1}\sum_{i=0}^{p}Y_{kpi}{t^{p+2i-1}\over \nu^{p}}+\dots \, ,
\ee
where the coefficients $Y_{kpi}$ can be calculated easily using a simple
computer program. In fact, Eq. \Ref{Ykpi} is the definition of the $Y_{kpi}$.
For Dirichlet ($\mu=+1$, $u=0$) and Neumann ($\mu=-1$, $u=0$) \bc they are
pure numbers; for Robin \bc ($\mu=-1$, $u\ne0$) they are polynomials in $u$.
In Eq. \Ref{Ykpi} the dots denote higher order terms which do not contribute
to the considered \hkksp

For the integration over $z$, we use the formula
\be\label{zint}\int_{0}^{\infty}dz \ z^{-1-2s} \
t^{-1+i}={\Gamma(-s)\Gamma(s+\frac{i-1}{2})\over 2\Gamma(\frac{i-1}{2})}.
\ee
One can easily check that the terms with $t^0$ from the asymptotic
expansions (\ref{uae1}) are cancelled after substitution inside the
brackets in $K_{D,R}$ (see eqs. (\ref{KDexp}) and (\ref{KRexp})).
This means that any new reflection contributes at least one power of $t$
to the integrand in (\ref{ank}). Therefore, according to (\ref{zint}),
only several first terms of the multiple reflection expansion
contribute to any given heat kernel coefficient. This explains the
finite range of the summations in \Ref{Ykpi}.

The sum over $l$ produces Hurwitz zeta functions.
When taking this into account, we obtain for the contribution of $k$ reflections to
$a_n$ (see \Ref{splitan}):
\beao a_{n}^{(k)}&=&\res{16\pi^{3/2}\over\Gamma(1-s)}
\sum_{p=k+1}^{2n+1}\sum_{i=0}^{p}Y_{kpi}{R^{2s}\mu^{k+1}\over 2^{k+1}}
\\ && ~~~~~~~~~~~~~~~~~~~~~~ \zeta(2s+p-2,\frac12){\Gamma(-s)\Gamma(s+i+\frac{p-1}{2})\over 2
  \Gamma(i+\frac{p-1}{2})} .
\eeao
The calculation of the residua can be carried out, again, using
standard computer algebra programs, which leads to the coefficients
$a_{n}^{(k)}$ in Eq. \Ref{splitan}.

\section*{Appendix 2}
Here we present the results for the \hkks corresponding to the
matching conditions in Sec. 4, in the cases of some higher
dimensional spaces. For $\beta=0$ we obtain, instead of Eq. \Ref{bn3}\\
in $D=4$ dimensions,
\beao
a_{1}  &=&  -2\,\left( \alpha - \gamma \right) \,{\pi }^2  \\
a_{3/2}&=&  \frac{1}{2}{\left( \alpha -\gamma\right) }^2\,{\pi}^{\frac{5}{2}}\\
a_{2}  &=&  - \frac{1}{6} \left( 2\,{\alpha}^3 - 105\,\gamma -
        2\,{\gamma}^3 \right) \,{\pi }^2  \\
a_{5/2}&=&   \frac{1} {64}\left( -3\,{\alpha}^2 + 4\,{\alpha}^4 -
         172\,\alpha\,\gamma -  4\,{\alpha}^3\,\gamma + 298\,{\gamma}^2 +
         3\,{\alpha}^2\,{\gamma}^2 - 10\,\alpha\,{\gamma}^3 +
         7\,{\gamma}^4 \right) \,{\pi }^{\frac{5}{2}}
\eeao
in $D=5$ dimensions,
\beao
a_{1}  &=& \frac{-8}{3}\,\left( \alpha - \gamma \right) \,{\pi }^2   \\
a_{3/2}&=& \frac{2}{3}\,{\left(\alpha -\gamma\right) }^2\,{\pi}^{\frac{5}{2}}\\
a_{2}  &=&-\frac{4}{45}\,\left( 5\,{\alpha}^3 - 465\,\gamma +
            5\,{\alpha}^2\,\gamma -  \alpha\,{\gamma}^2 -
            9\,{\gamma}^3 \right)\,{\pi }^2\\
a_{5/2}&=&   \frac{1}{12}\left( -2\,{\alpha}^2 + {\alpha}^4
          - 76\,\alpha\,\gamma + 128\,{\gamma}^2 - 4\,\alpha\,{\gamma}^3
          + 3\,{\gamma}^4 \right) \, {\pi }^{\frac{5}{2}}
\eeao
in $D=6$ dimensions,
\beao
a_{1}  &=&  - \left( \alpha - \gamma \right) \,{\pi }^3  \\
a_{3/2}&=&\frac{1}{4}{\left( \alpha -\gamma \right) }^2\,{\pi }^{\frac{7}{2}}\\
a_{2}  &=&   - \frac{1}{12} \left( 6\,\alpha + 2\,{\alpha}^3 - 297\,\gamma
             +4\,{\alpha}^2\,\gamma - 6\,{\gamma}^3 \right) \,{\pi }^3  \\
a_{5/2}&=&    \frac{1}{128}\left( {\alpha}^2 + 4\,{\alpha}^4
            - 508\,\alpha\,\gamma + 4\,{\alpha}^3\,\gamma + 802\,{\gamma}^2
            - {\alpha}^2\,{\gamma}^2 -  26\,\alpha\,{\gamma}^3
              + 19\,{\gamma}^4 \right) \,{\pi }^{\frac{7}{2}}
\eeao
and in $D=7$ dimensions,
\beao
a_{1}  &=& -\frac{16}{15}\,\left( \alpha - \gamma \right) \,{\pi }^3    \\
a_{3/2}&=&\frac4{15}\,{\left(\alpha-\gamma\right)}^2\,{\pi}^{\frac{7}{2}}\nn\\
a_{2}  &=&    -\frac{8}{225}\,\left( 5\,{\alpha}^3 - 1050\,\gamma
            + 15\,{\alpha}^2\,\gamma +3\,\alpha\,{\gamma}^2
               - 23\,{\gamma}^3 \right) \,{\pi }^3  \nn  \\
a_{5/2}&=&  \frac{1}{30}\left( -6\,{\alpha}^2 + {\alpha}^4
           - 172\,\alpha\,\gamma + 2\,{\alpha}^3\,\gamma + 280\,{\gamma}^2
            - 10\,\alpha\,{\gamma}^3
             +7\,{\gamma}^4 \right) \,{\pi }^{\frac{7}{2}}\,.
\eeao

\bigskip

For $\beta\ne0$ we obtain, instead of Eq. \Ref{D3},\\
in $D=4$ dimensions, \beao
a_{1}    &=& \frac{8}{\beta} \pi^{2}\\
a_{3/2}&=&\frac{1}{16}\,\left(
 9
+64\,\frac{\alpha}{\beta}
- 32\frac{\gamma}{\beta}
+\frac{128}{\beta^{2}}
\right){\pi }^{\frac 52} \\
a_{2}&=&
\frac{16}{3} \left(
 3\,\frac{\alpha}{\beta^{2}}
+\frac{4}{\beta^{3}}
\right) \pi^{2}\\
a_{5/2}&=&
\frac{1}{2048}\left(
     - 59
-       512\,\frac{\alpha}{\beta}
       - 224\,\frac{\gamma}{\beta}
-       512\,\frac{1}{\beta^2}
+       4096\,\frac{{\alpha}^2}{\beta^2}
+       2048\,\frac{\alpha\gamma}{\beta^{2}}
        + 512\,\frac{{\gamma}^2}{\beta^{2}}
\right.\nn\\&&\qquad\qquad\qquad      \left.
+       32768\,\frac{\alpha}{\beta^{3}}
+       8192\,\frac{\gamma}{\beta^{3}}
+32768 \frac{1}{\beta^{4}}
       \right) \pi^{\frac52}\label{D4}
\eeao
in $D=5$ dimensions,
\beao
 a_1&=&\frac{32}{3}\frac{\pi^{2}}{\beta}\nn \\
 a_{3/2}&=& \frac{2}{3}   \left(
 3
+ 2\,\frac{\alpha}{\beta}
-        2\,\frac{\gamma}{\beta}
+     \frac{4}{\beta^{2}}
\right){\pi }^{\frac{5}{2}} \nn\\
 a_2&=&\frac{16}{45}\left(
-     \frac{1}{\beta}
+ 60\,\frac{\alpha}{\beta^{2}}
- 20\,\frac{\gamma}{\beta^{2}}
+\frac{80}{\beta^{3}}
\right)\pi^2 \nn\\
 a_{5/2}&=&\frac{4}{3}\left(
-\frac{17}{240}
-       \frac{ \alpha}{\beta}
-        \frac{2}{\beta^{2}}
+ 2\,\frac{{\alpha}^2}{{\beta}^2}
+ 16\,\frac{\alpha}{\beta^{3}}
+  \frac{16}{\beta^{4}}
\right){\pi }^{\frac{5}{2}}
\label{D5}
\eeao
in $D=6$ dimensions,
\beao
 a_1&=&
\frac{4\,}{\beta}{\pi }^3\nn\\
 a_{3/2}&=&\frac{1}{32}\left(
       \frac{191}{3}
+        64\,\frac{\alpha}{\beta}
-      96\,\frac{\gamma}{\beta}
+  \frac{128}{\beta^{2}}
\right)\pi^{\frac 72}\nn\\
 a_2&=&\frac{2}{3}\left(
   \,\frac{3}{\beta}
+   12\,\frac{\alpha}{\beta^{2}}
-   8\frac{\gamma}{\beta^{2}}
+     \frac{16}{\beta^{3}}
\right){\pi }^3 \nn\\
 a_{5/2}&=&\frac{1}{12288}\left(
       103
+       512\,\frac{\alpha}{\beta}
      - 480\,\frac{\gamma}{\beta}
+       \frac{512}{\beta^2}
+       12288\,\frac{\alpha^2}{\beta^2}
-       6144\,\frac{\alpha\gamma}{\beta^{2}}
       - 512\,\frac{\gamma^2}{\beta^{2}}
\right.\nn\\ &&\qquad\qquad\qquad\qquad    \left.
+       98304\,\frac{\alpha}{\beta^{3}}
-       24576\,\frac{\gamma}{\beta^{3}}
+      \frac{98304}{\beta^{4}}
       \right){\pi }^{\frac 72} \label{D6}
\eeao
and in $D=7$ dimensions,
\beao
 a_1&=&\frac{64}{15}\frac{\pi ^3}{\beta}\nn\\
 a_{3/2}&=& 4 \left(
        10
+ 8\,\frac{\alpha}{\beta}
- 16\,\frac{\gamma}{\beta}
+    \frac{16}{\beta^{2}}
\right)\pi^{\frac72} \nn\\
 a_2&=&\frac{32}{225}\left(
      3
+ 60\,\frac{\alpha}{\beta^{2}}
- 60\,\frac{\gamma}{\beta^{2}}
+   \frac{80}{\beta^{3}}
\right)\pi^{3} \nn\\
 a_{5/2}&=&\frac{4}{15 }\left(
-         \frac{15}{8}
-        6\,\frac{\alpha}{\beta}
+        6\,\frac{\gamma}{\beta}
-       \frac{12}{\beta^2}
+        4\,\frac{\alpha^2}{\beta^2}
-        4\,\frac{\alpha\gamma}{\beta^{2}}
\right.\nn\\&&\qquad\qquad\qquad      \left.
+       32\,\frac{\alpha}{\beta^{3}}
       -16\,\frac{\gamma}{\beta^{3}}
+\frac{32}{\beta^{4}}
\right) {\pi }^{\frac{7}{2}} \,.
\label{D7}
\eeao

\bibliographystyle{unsrt}
\bibliography{mref,/home/bordag/Literatur/Bordag,/home/bordag/Literatur/Gilkey,/home/bordag/Literatur/libri,/home/bordag/Literatur/Kirsten,/home/bordag/Literatur/Osborn,/home/bordag/Literatur/articoli,/home/bordag/Literatur/Grosche}

\end{document}